\documentclass[doublecol]{epl2}
\usepackage{graphicx}
\usepackage{epstopdf}
\usepackage{epsfig}
\usepackage{amsfonts}
\usepackage{amsmath}
\title{Stall force of a cargo driven by $N$ interacting motor proteins}
\shorttitle{} 

\author{Deepak Bhat \inst{1} \and Manoj Gopalakrishnan \inst{2}}
\institute{
  \inst{1} International Centre for Theoretical Sciences, Tata Institute of Fundamental Research, Bangalore 560089, India\\
  \inst{2} Department of Physics, Indian Institute of Technology Madras, Chennai 600036, India
}

\abstract{
We study a generic one-dimensional model for an intracellular cargo driven by $N$ motor proteins against an external applied force. The model  includes motor-cargo and motor-motor interactions. The cargo motion is described by an over-damped Langevin equation, while motor dynamics is specified by hopping rates which follow a local detailed balance condition with respect to change in energy per hopping event. Based on this model, we show that the stall force, the mean external force corresponding to zero mean cargo velocity, is completely independent of the details of the interactions and is, therefore, always equal to the sum of the stall forces of the individual motors. This exact result is arrived on the basis of a simple assumption: the (macroscopic) state of stall of the cargo is analogous to a state of thermodynamic equilibrium, and is characterized by vanishing net probability current between any two microstates, with the latter specified by motor positions relative to the cargo. The corresponding probability distribution of the microstates under stall is also determined. These predictions are in complete agreement with numerical simulations, carried out using specific forms of interaction potentials. }

\begin{document}

\maketitle 

\section{Introduction}

Motor proteins, by consuming energy released during adenosine triphosphate (ATP) hydrolysis, pull and transport 
intracellular organelles on the cytoskeletal filaments \cite{Debashish}. Their activity is crucial for many biological 
processes such as axonal transport of mitochondria in neurons \cite{Goldstein}, lipid droplet transport in {\it 
Drosophila} \cite{Gross2}, pigment granule transport in {\it Xenopus laevis} \cite{Nascimento} and so on. Broadly, 
spatial organisation of different biomolecular structures inside eukaryotic cells is made possible due to the cargo 
transport by motor proteins. When motor proteins stop functioning, major biological activities slow down and often such 
disruption in transport is lethal for the cell \cite{Mandelkow}. The biophysics of motor proteins has been an active 
interdisciplinary area of research for many decades now, specific areas of interest including mechanochemical coupling, 
free energy transduction, thermodynamic efficiency, force generation, directed motion and many other collective 
phenomena. 

Typically, a motor protein on a cytoskeletal filament moves with a speed of hundreds of nanometres per second and exert forces of a few piconewtons on a cargo. The maximum force with which a motor protein pulls an organelle on the filament is called its \emph{stall force} \cite{Debashish,Fisher}. In {\it in vitro} optical trap experiments, stall force is determined as the opposing external force at which average velocity of the motor-driven cargo becomes zero \cite{Block}. Collective force generation by teams of motor proteins has been a subject of interest in the recent past \cite{Kunwar2,Soppina,Hendricks,Deepak1,Arpan}. When a cargo is driven by multiple motors, it is generally assumed that the stall force scales linearly with the number of motors, and hence the measured stall force is also used to determine the number of cargo-bound motors in experiments \cite{Vershinin,Soppina,Arpan}. In reality, motor-cargo as well as motor-motor interactions \cite{Rogers} may be present, and it is unclear how these interactions within the motor-cargo complex affect the stall force. In Camp\`as et al. \cite{Campas}, it was shown that direct motor-motor interaction captured effectively via a simplified model influences the force-velocity behaviour as well as the stall force of a multiple-motor assembly \cite{Campas}. The role of motor-cargo interaction has been the subject of many earlier experimental \cite{Rogers} and theoretical \cite{Kunwar,Kunwar2, Bouzat,Bouzat2,Deepak} studies. In our recent work \cite{Deepak} on the dynamics of a cargo elastically coupled to an arbitrary number of motor proteins, numerical simulations showed that, while the average velocity and diffusion coefficient of the cargo is a decreasing function of the stiffness of the motor-cargo linker, the stall force is completely independent of it, and is equal to the sum of stall forces of the individual motors. This observation inspired the present study; here we seek to identify the underlying physical principle behind this simple and remarkable property of a multiple-motor assembly. 

In this letter, using a generic model consistent with basic thermodynamic principles, we investigate the stall force of a cargo pulled by \emph{interacting} motor proteins, involving both direct motor-motor as well as motor-cargo interactions. We argue that the state of stall of the cargo is analogous to thermodynamic equilibrium, with net zero current within any pair of configurational states specified by the positions of each motor relative to the cargo. Under this single assumption, it follows, as an exact result, that the mean external force under stall depends solely on the intrinsic properties of the individual motors and neither motor-motor nor motor-cargo interaction plays any role in determining it. The generality of the result explains why the standard experimental  practice of estimation of the number of cargo-bound motors from measurement of stall force is robust when all attached motors are of the same directionality. Numerical simulations, with various choices for interaction potentials, completely support our theory. 

{\it Model:} We consider a cargo pulled by $N$ motor proteins of combined plus and minus directionalities (see fig.1). Each motor protein is assumed to move in a random sequence of forward and backward jumps of fixed length  $\ell$ on a microtubule filament with rates $w^{+}_i$ and $w^{-}_i$ respectively ($1\leq i\leq N$). A single motor, on an average, moves in the plus direction when $w^{+}_i>w^{-}_i$ and minus direction when $w^{+}_i <  w^{-}_i$. If $x_0$ and $x_i$ $(1\leq i\leq N)$ are the instantaneous positions of the cargo and the i'th motor on the filament respectively, then $\Delta_i\equiv x_i-x_0$ is the instantaneous separation between them. For the sake of later convenience, we introduce compact vector notations ${\bf x} \equiv \{x_0,x_1,...x_N\}$ and ${\bf \Delta} \equiv\{\Delta_1, \Delta_2, ...\Delta_N\}$ for position and stretch variables respectively. Let $U({\bf x})$ be the total energy of the motor-cargo system in a configuration ${\bf x}$ such that
\begin{eqnarray}
U({\bf x})=U_{\rm mc}({\bf x})+U_{\rm mm}({\bf x}) + U_{\rm e} (x_0). 
\label{e0}
\end{eqnarray}
Here, $U_{\rm mc}({\bf x})=\sum^{N}_{i=1} u_i(\Delta_i)$ is the total motor-cargo interaction energy, with $u_i(\Delta_i)$ denoting the same between the $i$'th motor and the cargo, which is assumed to depend only on the motor-cargo separation $\Delta_i$. $U_{\rm mm}({\bf x})=\sum^{N}_{i,j> i} v_{ij}(\Delta_{ij})$ is the net motor-motor direct interaction energy which is assumed to be short-range (e.g. hard-core or soft repulsion), and depending only on the mutual separations $\Delta_{ij}=\Delta_j-\Delta_i$. In {\it in vitro} experiments, a cargo is usually brought under influence of an optical tweezer either to confine it or to a exert controlled force on it; this external background potential which influences only the cargo is represented by the term $U_{e}(x_0)$ in eq.\ref{e0}. The over-damped Langevin equation for the motion of the cargo is written as \cite{Bouzat,Deepak}:
\begin{eqnarray}
\dot{x}_0= -\gamma^{-1}\partial_{x_0} \left[U_{\rm mc}+U_{\rm e}\right] +\zeta(t),
\label{e1}
\end{eqnarray} 
where $\gamma$ is the drag coefficient, $\zeta(t)$ is Gaussian white noise: $\langle \zeta(t) \rangle=0$ and $\langle \zeta(t) \zeta(t^{\prime})  \rangle= 2 D \delta(t-t^{\prime})$, with $D=k_BT/\gamma$ following fluctuation-dissipation theorem. The mean velocity of the cargo $V_0\equiv \langle \dot{x}_0\rangle$ is obtained by averaging eq.\ref{e1} over ${\bf x}$. Throughout this paper, we shall use $\partial_{y}$ to represent the partial derivative with respect to any variable $y$.
\begin{figure}
\centering
\includegraphics[angle=-90,width=7cm,keepaspectratio=true]{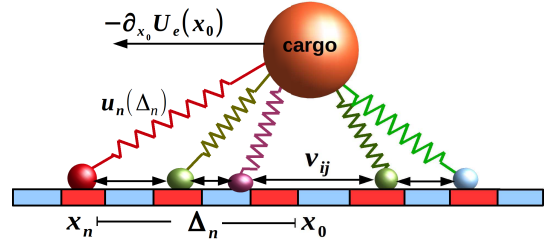}
\caption{\label{f1} The figure depicts a cargo at position $x_0$ interacting with filament-bound 
multiple motor proteins at $x_n$ ($1\leq n\leq N$). $u_n(\Delta_n)$ is the motor-cargo interaction potential and 
$-\partial_{x_0}U_{\rm e}(x_0)$ is the external force. Two sided arrows indicate motor-motor direct interaction 
$v_{ij}(\Delta_{ij})$ between the $i$'th and $j$'th motor.} 
\end{figure}

The interactions within the motor-cargo system modifies the hopping rates of the individual motors. Let us denote by $W_i^{+}({\bf \Delta})$ and $W_i^{-}({\bf \Delta})$ the single motor hopping rates for the forward ($x_i\to x_i+\ell$) and backward ($x_i\to x_i-\ell$) transitions respectively. Now, we define local energy differences
\begin{eqnarray}
\epsilon^{\pm}_i=\pm (\mathbb{E}_i^{\pm} -1) \left[u_i(\Delta_i) + \sum_{j\neq i} v_{ij}(\Delta_{ij}) \right]
\label{e2}
\end{eqnarray}
for a certain motor $i$ at position $x_i$, corresponding to a single hop to its right or left, where $\mathbb{E}_i^{+}$ 
and $\mathbb{E}_i^{-}$ are a set of $N$ raising and lowering operators, defined through the relations 
$\mathbb{E}_i^{\pm} g(x_0..x_i..x_N)=g(x_0..x_i\pm \ell..x_N)$. We assume further that the forward and backward hopping 
events are tightly coupled to ATP hydrolysis and the corresponding rates follow a local detailed balance condition 
(LDBC)\cite{Zimmermann}, i.e., 
\begin{eqnarray}
\frac{W^+_i({\bf \Delta})}{\mathbb{E}_i^{+} W^-_{i}({\bf \Delta})} = \exp\left[ \beta(\mu_i - \epsilon^{+}_i) \right]
\label{e3}
\end{eqnarray}
where $\beta=(k_B T)^{-1}$; $\mu_i=k_BT \log(w^{+}_i/w^{-}_i)$ is the amount of free energy made available by ATP 
hydrolysis to the $i$'th motor, per unit forward step. This is an important assumption which has been employed in many 
earlier studies \cite{Zimmermann} and in this paper, it serves as key ingredient in determining the stall force.
Although an explicit form for these rates is not essential for further analysis, while carrying out numerical 
simulations, we use rates $W^{+}_{i}({\bf \Delta}) =w^{+}_i \exp\left[-\beta \epsilon_i^{+} \theta_i \right] $ and $ 
W^-_{i}({\bf \Delta})=w^{-}_i \exp\left[\beta \epsilon_i^{-} (1-\theta_i) \right]$ where $0 \leq \theta_i \leq 1$  is a 
dimensionless number that  characterizes the asymmetry of the energy landscape for the motion of the motor 
\cite{Zimmermann}.

Let us define $P({\bf x};t)$ as the probability density for ${\bf x}$, which is normalized as $\sum_{i}\sum_{x_i}\int dx_0 P({\bf x};t)=1$, where $1\leq i\leq N$ and the summation involves the discrete motor positions on the filament. The dynamics of $P({\bf x};t)$ is described by the master equation
\begin{eqnarray}
\partial_{t} {P({\bf x};t)}  = \sum_{i=1}^{N} \left[ (\mathbb{E}_i^{+} - 1)W_i^{-}P+(\mathbb{E}_i^{-} - 1)W_i^{+}P \right] \nonumber\\ + \partial_{x_0} \left[ \gamma^{-1}\partial_{x_0}\left({\rm U_{mc}+U_{\rm e}}\right) P+ 
D \partial_{x_0} P \right] 
\label{e4}
\end{eqnarray}
It is pertinent to note that $P({\bf x};t)$ does not necessarily become time-independent in the long-time limit (e.g. when $U_{e}=-f x_0$ with constant $f$). However, the distribution 
\begin{eqnarray}
\psi({\bf \Delta};t)=  \sum^{N}_{i=1} \sum_{x_i} \int dx_0 P({\bf x};t) \prod^{N}_{i=1}\delta[\Delta_i-(x_i-x_0)]
\label{e5}
\end{eqnarray}
for the stretch variables ${\bf \Delta}$ is expected to be, because, as motor proteins pull the cargo, viscous drag offers an opposing force to them thereby increasing the motor-cargo separation. But, due to increase in the stretching energy, the motor-cargo separation cannot increase arbitrarily. Consequently, the average separation between any motor and the cargo is bounded in the long time limit. Therefore, it is convenient to describe the motor-cargo dynamics in terms of $\psi({\bf \Delta};t)$, which is normalized as  $\int \psi({\bf \Delta} ;t )d{\bf \Delta} =1$, where $d{\bf \Delta} \equiv \prod_{i=1}^{N}d\Delta_i$. We next transform the operators in eq.\ref{e4}; while $\mathbb{E}_i^{\pm} \Phi(x_0,\Delta_1,...\Delta_i,..\Delta_N) = \Phi(x_0,\Delta_1, ..\Delta_i\pm \ell,..\Delta_N)$, $\partial_{x_0} \to \partial_{x_0}- \sum^{N}_{i=1} \partial_{\Delta_i}$. With these changes, the master equation in eq.\ref{e4} takes the form
\begin{eqnarray}
\partial_{t} \psi({\bf \Delta};t) =  - \sum_{i=1}^{N}  \left( J^{\mathcal M}_{i+}+ J^{\mathcal M}_{i-} \right) - \sum_{i=1}^{N} \partial_{\Delta_i} (J^{\mathcal C}) .
\label{e7}
\end{eqnarray}
Here
\begin{eqnarray}
J^{\mathcal C} =- \gamma^{-1}\left[ \sum^{N}_{j=1}  (\partial_{\Delta_j} u_j)  \psi + F_e({\bf \Delta}) \right]- D\sum^{N}_{j=1}  \partial_{\Delta_j} \psi
\label{e8}
\end{eqnarray}
is the probability current due to the cargo dynamics; 
\begin{eqnarray}
F_e({\bf \Delta})= - \sum^{N}_{i=1} \sum_{x_i}\int dx_0 \partial_{x_0}U_{\rm e}P({\bf x};t) \prod^{N}_{i=1}\delta[\Delta_i-(x_i-x_0)], 
\label{e9}
\end{eqnarray}
from which the average external force $f_e$ on the cargo  is determined as 
\begin{eqnarray}
f_e=\int F_e({\bf \Delta})  d{\bf \Delta}. 
\label{e9+}
\end{eqnarray}
In particular, if the potential is $U_{\rm e}=-fx_0$ (constant external force $f$), then $f_e=f$, on the other hand if $U_{\rm e} = (\kappa_{t} /2) x^2_0$ (cargo trapped by an optical tweezer with trap stiffness $\kappa_{t}$), $f_e= -\kappa_{t} \langle x_0 \rangle$. Further, 
\begin{eqnarray}
 J^{\mathcal M}_{i\pm}= W_i^{\pm} ({\bf \Delta})\psi({\bf \Delta}) -  \mathbb{E}_i^{\pm}\left[W_i^{\mp}({\bf \Delta}) \psi({\bf \Delta}) \right]
\label{e10}
\end{eqnarray}
denote the {\it net} probability currents associated with the $\Delta_i$ to $\Delta_i \pm \ell$ hopping events of the $i$'th motor. Notice that eq.\ref{e7} comprises both discrete dynamics due to hopping of motor proteins and continuous dynamics due to the cargo motion. As $t\rightarrow \infty$, the probability distribution function $\psi({\bf \Delta},t)$ becomes stationary: define $\psi({\bf \Delta})\equiv\psi({\bf \Delta};t \rightarrow \infty$). As a result, the average separations $\langle\Delta_i\rangle=\int \Delta_i \psi( {\bf \Delta} ;t)d{\bf \Delta}$ assume constant values; $d\langle\Delta_i\rangle/dt=0$. Employing this in eq.\ref{e7}, we see that for all $1\leq i\leq N$, $V_i=V_0$ in steady state, where
\begin{equation}
V_0=\gamma^{-1}\left
[\sum^{N}_{j=1} \langle\partial_{\Delta_j} u_j \rangle + f_e\right]
\label{e10+}
\end{equation}
is the mean cargo velocity (from eq.\ref{e1}), $V_i\equiv \ell\left[\langle W_i^{+} \rangle - \langle W_i^{-}\rangle \right]$ give the mean motor velocities.

Among the plethora of steady states associated with a given master equation, the equilibrium state is special: it satisfies the condition of detailed balance and hence vanishing probability current between any pair of states \cite{VanKampen}. Does such a state exist in the present system ? To explore this possibility, we impose the conditions
\begin{equation}
J^{\mathcal C}=0~~~;~~~J^{\mathcal M}_{i\pm}=0.
\label{e11}
\end{equation}
Integrating the first of eq.\ref{e11} over ${\bf\Delta}$ yields $V_0=0$, after using eq.\ref{e10+}; therefore, when $J^{C}=0$, the cargo is under stall. The mean external force at stall is  given by $f_e=f_s^N$, where
\begin{equation}
f_s^N = - \sum^{N}_{j=1} \langle \partial_{\Delta_j} u_j \rangle,
\label{e11+}
\end{equation}
the evaluation of which is our next task. The second part of eq.\ref{e11}, i.e., $J_{i\pm}^{\mathcal M}=0$, implies that $W_i^{-}({\bf\Delta}) \psi({\bf \Delta})  = \mathbb{E}_i^{-}\left[ W_i^+({\bf \Delta}) \psi({\bf \Delta})\right]$ . Using the LDBC in eq.\ref{e3}, we recursively solve for $\psi({\bf \Delta})$ to find 
\begin{eqnarray}
 \psi({\bf \Delta}) = \psi({\bf 0}) \exp\left[- \beta \sum^{N}_{i=1}\left( u_i + \sum_{j>i} v_{ij}-
 \frac{\mu_i}{\ell}\Delta_i\right) \right].
 \label{e13}
\end{eqnarray}
The right hand side of eq.\ref{e11+} is easily evaluated from the distribution in eq.\ref{e13} by noting that $\int \partial_{\Delta_i}\psi({\bf\Delta})d\Delta_i=0$ since we require $\psi({\bf\Delta})=0$ at $\Delta_i=\pm\infty$ for any $i\in [1,N]$. Upon taking the derivative, we find, for each $i$, 
\begin{equation}
\langle\partial_{\Delta_i} u_i\rangle + \sum_{j\neq i}\langle \partial_{\Delta_i} v_{ij}\rangle-\ell^{-1}\mu_i=0.
\label{e13a}
\end{equation}
After summing eq.\ref{e13a} over all $i\in[1,N]$, the second term vanishes (since $v_{ij}$ are functions of the separations $\Delta_{ij}$, $\sum_{i,j\neq i}\partial_{\Delta_i} v_{ij}=0$ identically). It follows that $\sum_{i} \langle \partial_{\Delta_i}u_i\rangle=\ell^{-1} \sum_{i} \mu_i$, and thus, the final expression for the stall force in eq.\ref{e11+} is given by
\begin{equation}
f_s^N =\frac{1}{\beta \ell}\sum^{N}_{i=1}\log\left(\frac{w^{-}_i}{w^{+}_i}\right). 
\label{e14}
\end{equation}

It is remarkable that the expression for stall force in this model, as given by eq.\ref{e14}, is simply the sum of stall forces of the individual motors \cite{Fisher}. The average cargo velocity and the force experienced by individual motors are known to depend significantly on the motor-cargo interaction in sub-stall conditions \cite{Deepak}, but the stall force itself turns out to be completely independent of it, in agreement with previous simulations \cite{Deepak}. It should also be mentioned that the earlier study by Camp\'as et al.\cite{Campas} did suggest dependence of the stall force on motor-motor interaction. However, their model has two major differences with our model: in \cite{Campas}, (i) the external force is experienced only by the leading motor, a constraint not present in our model, and  (ii) the modification in hopping rates of a single motor due to the presence of neighbouring motors is not derived based on an underlying interaction potential, while in our model, this aspect is taken care of explicitly through the LDBC in eq.\ref{e3}. 

The result in eq.\ref{e14} has several important implications. In particular, for $K$ identical motors ($w^{+}_i=w^{+}, w^{-}_i=w^{-}$), the combined stall force $f_s^{K}=Kf_s^{1}$, while for a cargo pulled by two opposing teams of identical motors of $K$ and $D$ members each, $f_s^{K,D}= Kf_s^{+} + Df_s^{-}$ where $f_s^{\pm}$ gives the stall force of a single plus/minus moving motor respectively (such that $f^+_s<0$ and $f^-_s>0$). Tug-of-wars between two dissimilar motor teams (dyneins and kinesins) have been observed in experiments directly \cite{Soppina} as well as indirectly \cite{Hendricks2}. Here, the stall forces of single dynein and kinesin were measured to be 1.1pN and $-5.5$ pN respectively. Recently, reduction in the stall force of kinesin due to binding of dynein has been reported \cite{Blechm}. Eq.\ref{e14} suggests that, in such cases, the net stall force is the sum of the combined stall forces of individual teams, and is independent of the nature of the motor-cargo as well as motor-motor coupling. 
\begin{table}
\centering
\begin{tabular}{|c|c|c|c|c|c|}
\hline
$\beta$               &      $\gamma$     &$w^+$           &$w^-$           &$\ell$    &$\theta$\\
${\rm pN^{-1}nm^{-1}}$&${\rm pNs~nm^{-1}}$&${\rm s^{-1}}$&${\rm s^{-1}}$&${\rm nm}$&        \\\hline
$0.2433$              &$9.42\times10^{-4}$&$125.0028$    &$0.0028$      &$8$       &$0.1$   \\\hline
\end{tabular} 
\caption{A list of parameters used (see \cite{Deepak} for details) in our calculations and numerical simulations are shown here.}
\label{t1}
\end{table}

To verify our results in computer simulations, we considered specific forms for $u_i(\Delta_i)$ and $v_{ij}(\Delta_{ij})$. A few experiments have pointed out the possibility of an-harmonic motor-cargo interaction \cite{Coppin}. Bearing this in mind, we choose $u_i(\Delta_i)=(\kappa_i/2) \Delta^2_i + (\lambda_i/4) \Delta^4_i$ ($\kappa_i,\lambda_i \geq 0$). For motor-motor interaction, nearest neighbour soft repulsive interaction of the form $v_{ij}(\Delta_{ij}) =A \left[ \delta_{ji-1} \exp(-\Delta_{ji}/b)+\delta_{ji+1}\exp(-\Delta_{ij}/b) \right]$ is used, where $A>0$ and $b>0$. Further, we set $\kappa_i=\kappa$, $\lambda_i=\lambda$ in all simulations for the sake of simplicity.  For a set of $K$ identical motors, we use  $w_i^{\pm}=w^{\pm},~ \theta_i=\theta$ for $1\leq i\leq K$. For set of $N=K+D$ (K plus moving and D 
minus moving) motors, we choose $w_i^{\pm}=w^{\pm},~ \theta_i=\theta,$ for $1\leq i\leq K$ and $w_i^{\pm}=w^{\mp},~ \theta_i=1-\theta,$ for $K+1\leq i\leq N$. A list of the parameters used in simulations are tabulated in tab.\ref{t1}; further details of simulations are to be found in \cite{Deepak}. 

\begin{figure}
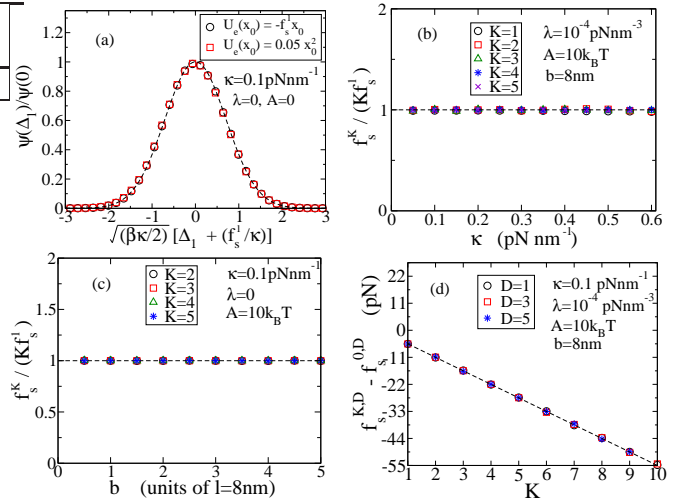

\centering
\resizebox{0.23\textwidth}{!}{\includegraphics{fig2a.eps}}~
\resizebox{0.23\textwidth}{!}{\includegraphics{fig2b.eps}}\\
\resizebox{0.23\textwidth}{!}{\includegraphics{fig2c.eps}}~~
\resizebox{0.23\textwidth}{!}{\includegraphics{fig2d.eps}}\\
\caption{\label{f3}  In (a), (rescaled) $\psi(\Delta_1)$ obtained in simulations(symbols), when the cargo is driven by a single motor $(K=1)$, is compared with eq.\ref{e13} (line) for two forms of $U_{\rm e}(x_0)$. The scaled stall force $f_s^K/(K f_s^1)$ is shown as a function of $\kappa$ in (b) and as a function of $b$ in (c) for $K$ identical plus-moving motor proteins. In (d), the shifted stall force $f^{K,D}_s - f^{0,D}_s$ of a cargo driven by two teams of antisymmetric motors is shown as a function of $K$. The values of varying parameters used in the simulations are mentioned in each plot and the (fixed) values of the remaining parameters are given in tab.\ref{t1}. The lines in (b)-(d), from eq.\ref{e14}, agree well with simulation results, shown with symbols.}
\end{figure}

For a cargo driven by a single motor ($K=1$), $\psi(\Delta_1)$ obtained in computer simulations (symbols), for $U_{\rm e}(x_0)= -f^1_s x_0$  (corresponds to a constant force equal to the stall force of a single motor) and $U_{\rm e}(x_0)=0.05 x^2_0$ (akin to an optical trap) is compared with eq.\ref{e13} for $\lambda=A=0$ in fig.\ref{f3}(a). The agreement between theory and simulation for both forms $U_{\rm e}(x_0)$ supports our conjecture that the conditions in eq.\ref{e11} accurately represent the stall state. Next, the stall force determined in simulations is given as a function of $\kappa$ and $b$ in fig.\ref{f3}(b) and (c) respectively for different numbers $K$ of identical motors. Further, to verify the generality of the result in eq.\ref{e14}, stall force is determined in simulations for a cargo pulled by two opposing teams of motors; the result is shown in fig.\ref{f3}(d) as a function of the number of plus moving motors $K$, with number ($D$) of minus-directed motors held fixed. In all our simulations, we find that the stall force is completely independent of the parameters that characterize the motor-motor or motor-cargo interaction, and satisfies eq.\ref{e14}. 

\emph{Summary:} Stall force of a multiple motor-driven cargo is commonly studied in {\it in vitro} experiments and is a useful quantity to estimate the number of cargo-bound motors \cite{Block,Vershinin,Soppina,Arpan}. In general, details of the interactions within the motor-cargo complex are not taken into account in estimating these numbers.  Here, we investigated the stall force of a cargo coupled to $N$ mutually interacting motor proteins. By drawing an analogy between the state of stall and thermodynamic equilibrium characterized by vanishing currents, we showed that the stall force is independent of both motor-motor and motor-cargo interactions. In particular, eq.\ref{e14} guarantees that this net force is the same for a plastic bead in an {\it in vitro} experiment as it is for a flexible intracellular cargo like an endosome, as in \cite{Soppina}, as long as the numbers of motors remain the same. This is the most important implication of our result. Many experiments with motor proteins carried out in a variety of cells, both {\it in vitro} \cite{Kunwar,Soppina,Shubeita} and {\it in vivo} \cite{Arpan,Blechm} give evidence for a simple linear scaling of the stall force with the number of motors. In our view, eq.\ref{e14} captures the fundamental principle underlying these observations. Finally, motor detachment from and reattachment to the filament was not included in the present study, although, in general, both are  important in cargo transport \cite{Gross}. This is because stall force measurements are generally done over small enough time intervals such that the probability of  motor detachment is negligible. Nevertheless, inclusion of detachment is important in understanding other properties of the stall state, for instance, residence time of a cargo in an optical trap or the duration of tug-of-war between two teams of motors. A suitable extension the present model including motor detachment is on the anvil.

\acknowledgments
DB thanks Udo Seifert and Vijaykumar Krishnamurthy for stimulating discussions. We acknowledge useful conversations with Ronojoy Adhikari.

\end{document}